\documentstyle[aps,floats,12pt]{revtex}

\input epsf
\def\BNL {Department of Physics, Brookhaven National Laboratory, Upton, NY 11973\\}
\newcommand{\LL}{\left\langle}
\newcommand{\RR}{\right\rangle}

\newcommand{\BE}{\begin{equation}}
\newcommand{\EE}{\end{equation}}
\newcommand{\BEA}{\begin{eqnarray}}
\newcommand{\EEA}{\end{eqnarray}}
\newcommand{\EL}{\nonumber\\}

\newcommand{\etal}{{\em et al.\ }}

\newcommand{\ie}{{\em i.e.\ }}
\newcommand{\vs}{{\em vs.\ }}

\newcommand{\gbeta}{6/g^2}
\newcommand{\CS}{{\rm SU(N_f)_L\times SU(N_f)_R}}
\newcommand{\MLL}{\LL \bar P |O_{LL}| P\RR}
\newcommand{\MK}{\LL \bar K |O_{LL}| K\RR}

\def\simge{
    \mathrel{\rlap{\raise 0.511ex
        \hbox{$>$}}{\lower 0.511ex \hbox{$\sim$}}}}
\def\simle{
    \mathrel{\rlap{\raise 0.511ex
        \hbox{$<$}}{\lower 0.511ex \hbox{$\sim$}}}}

\begin{document}

\title{ Domain wall quarks and kaon weak matrix elements } 
\author{T. Blum and A. Soni}
\address{ \BNL }
\date{\today}

\maketitle
\vskip .25in
\begin{abstract}
We present lattice calculations of kaon matrix elements 
with domain wall fermions. 
Using lattices with $\gbeta=5.85$, 6.0, and 6.3,
we estimate $B_K(\mu\approx 2\,{\rm GeV})=0.628(47)$ in quenched QCD which 
is consistent
with
previous calculations. At $\gbeta=6.0$ and 5.85 we find
the ratio $f_K/m_\rho$ in
agreement with the experimental value, within errors.
These results support expectations that $O(a)$ errors are
exponentially suppressed in low energy ($E\ll a^{-1}$) observables, and 
indicate
that domain wall fermions have good scaling behavior at relatively
strong couplings. 
We also demonstrate that the axial current
numerically satisfies the lattice analog of the usual 
continuum axial Ward identity.
\end{abstract}
\vskip 0.5in

While lattice gauge theory has made significant progress in
addressing the outstanding challenge of calculating hadronic observables
from first principles, a basic feature of the strong interactions has been 
missing in these calculations, the $\CS$ chiral flavor symmetry of the
light quarks which is broken explicitly by present lattice discretizations
of continuum QCD. 
We recently reported\cite{US} on calculations
using a new discretization for simulations
of QCD, domain wall fermions
(DWF)~\cite{KAPLAN,SHAMIR}, which preserve chiral symmetry on the lattice
in the limit of an infinite extra 5th dimension.
There it was demonstrated
that DWF exhibit remarkable chiral behavior\cite{US} even at relatively large
lattice spacing and modest extent of the fifth dimension.
Here we give further results using DWF which are of
direct phenomenological interest.
 
In addition to retaining chiral symmetry, DWF are also
``improved'' in another important way. 
In the limit that the number of sites in the extra dimension, $N_s$, goes 
to infinity, the leading discretization error in the
effective four dimensional action for the light degrees of freedom goes like
$O(a^2)$, unlike the case for ordinary Wilson fermions, for which
the errors are $O(a)$, $a$ being the lattice spacing.
This theoretical dependence
is deduced from the fact that the only operators available to
cancel $O(a)$ errors in the effective action
are not chirally symmetric; thus no $O(a)$ errors exist
in the low energy theory.
For finite $N_s$, $O(a)$ corrections are expected to be exponentially 
suppressed with the size of the extra fifth dimension. 
Our calculations for $B_K$ show 
a weak dependence on $a$ that is easily fit to an $a^2$ ansatz.
At $\beta=6.0$ ($\beta\equiv \gbeta$) the lattice spacings
determined from $m_\rho$ and $f_\pi$ agree within less than five percent.
This improved scaling behavior is plausible in light of the fact that
DWF retain an important continuum symmetry at non-zero
lattice spacing.
 

As in our previous paper, we
use the boundary fermion variant of DWF\cite{KAPLAN} developed by Shamir\cite{SHAMIR}.
The DWF action is essentially a five dimensional analog of the ordinary
Wilson fermion action with two key differences: (1) the relative sign 
between the Wilson term and the (five dimensional) Dirac mass, $M$, 
is opposite to the usual convention. This leads to
the appearance of massless chiral modes on the boundaries of the 
fifth dimension, a
left handed fermion on one wall and a right handed  one on the other. (2) The
layers $s=0$ and $s=N_s-1$ (s denotes the coordinate in the extra dimension)
are coupled with strength $-m$ $(m\ge 0)$. 
Neglecting exponentially small corrections, in Ref.\cite{SHAMIR} it was shown that the
parameter $m$ is (proportional to) the mass of the light four dimensional 
quark which is assembled from the two chiral modes, $m_q=m M(2-M)$. 
The chiral limit is $N_s \to \infty$ and $m\to 0$,
which requires no fine tuning unlike ordinary Wilson fermions.

Recently the exponentially small corrections to the quark mass have been
given at tree level~\cite{VRANAS}, $m_q=M(2-M)(m + \{1-M + O(p^2)\}^{N_s})$. 
In the presence of interactions $M$ is renormalized additively just
like ordinary Wilson fermions which also acquire a ``mass term" proportional to $p^2$. 
Perturbatively at one loop the main effect of the interactions is 
the replacement
$M\to M + g^2 \Sigma(p)$\cite{SHAMIR} where $\Sigma(p)$ is the 
quark self-energy.
Thus the chiral limit still holds with 
the replacement $(1-M)^{N_s}\to (M_c-M)^{N_s}$.
This is analogous to the renormalization of the
critical hopping parameter from its tree level value of 1/8
in the case of ordinary Wilson fermions. 
Of course, the crucial difference is that for DWF
the additive corrections are exponentially suppressed.
In our original study we found $M_c \approx 1.7$ 
for non-perturbative couplings corresponding
to quenched simulations at $\beta\sim 6.0$, 
which also agrees roughly with a simple
mean field argument\cite{US}.

For QCD, the DWF are gauged in the ordinary four dimensions only, and the
left and right handed modes couple equally
to the gauge field. Thus the five dimensional theory gives rise to
a low energy effective theory ($E<<a^{-1}$) describing interacting 
vector quarks in four dimensions whose right and left handed 
components are localized around $s=0$ and $s=N_s-1$, respectively. 

In Ref.~\cite{SHAMIRandFURMAN} it was shown that operators constructed
from the quark fields formed by the chiral modes on each wall
satisfy the following four dimensional chiral Ward identities(CWI).
\BEA
\Delta_\mu \LL A^a_\mu(x) O(y_1, y_2,...)\RR &=& 
2 m \LL J_5^a(x)O(y_1, y_2,...)\RR \EL
&+& 2 \LL J_{5q}^a(x)O(y_1, y_2,...)\RR\EL 
&+& i \LL \delta_A^a O(y_1, y_2,...)\RR,
\label{cwi}
\EEA
which result from demanding invariance of $\LL O(y_1, y_2,...)\RR$
under an infinitesimal axial transformation $\delta_A^a$.
Here $A^a_\mu$ is the four dimensional partially conserved axial current (PCAC) 
constructed from a sum of five dimensional fields over all slices in the extra
dimension\cite{SHAMIRandFURMAN}.
The operators $O(y_1, y_2,...)$ and the pseudoscalar density
$J^a_5(x)$ are constructed from four dimensional
quark fields using the chiral modes on each boundary, 
\BEA
q(x) &=& \frac{1+\gamma_5}{2}\psi(x,0)+\frac{1-\gamma_5}{2}\psi(x,N_s-1), \EL
\bar q(x) &=& \bar\psi(x,N_s-1)\frac{1+\gamma_5}{2}+\bar\psi(x,0)\frac{1-\gamma_5}{2}.
\label{quarks}
\EEA
In Eq.~\ref{cwi}, $J^a_{5q}$ is an anomalous pseudoscalar density
that results from the non-invariance of the action under 
the axial transformation for finite $N_s$. 
For flavor non-singlet currents, this contribution to the 
r.h.s. of Eq.~\ref{cwi} vanishes identically in the limit 
$N_s\to \infty$\cite{SHAMIRandFURMAN}, and we are left with
the continuum-like relations.
Below we demonstrate explicitly that at $\beta=6.0$
Eq.~\ref{cwi} is satisfied
for the usual PCAC Ward identity,
$O(y_1, y_2,...)=J_5$, and the anomalous contribution is
small for $N_s=10$ and reduces further by more than a factor
of 2.5 as $N_s$ is increased to 14.

To obtain $B_K$ we need the matrix element of the $\Delta s=2$
four quark operator that governs $K-\bar K$ mixing,
$O(\mu)_{LL} = [\bar s \gamma_\nu ( 1- \gamma_5) d]^2$,
which depends on the energy scale $\mu$. On the lattice and using DWF, 
$O_{LL}$ is a simple transcription of the above using the quark fields 
in Eq.~\ref{quarks}. Sandwiching $O_{LL}$ between
$K$ and $\bar K$ states and taking the ratio with its value in vacuum
saturation yields $B_K$.


\begin{table}[hbt]
\caption{Summary of simulation parameters.
``size" is the number of spatial sites times the 
temporal extent times $N_s$. $M$ is the five dimensional Dirac
fermion mass, and $m$ is the coupling between layers $s=0$ and 
$N_s-1$. The number in parentheses is the number of configurations
used at each value of $m$.}
\begin{tabular}{cccc}
$\beta$ & size & $M$ & $m$($\#$ conf) \\
\hline
5.85& $16^3\times 32\times 14$& 1.7& 0.075(34), 0.05(24)\\
6.0 & $16^3\times 32\times 10$& 1.7& 0.075(36), 0.05(39), 0.025 (17)\\ 
6.3 & $24^3\times 60\times 10$& 1.5& 0.075(11), 0.05(14)\\
\end{tabular}
\label{RUNTABLE}
\end{table}

In order to investigate the continuum limit of quenched QCD with DWF,
we have carried out simulations at gauge couplings $\beta=5.85$, 6.0, and
6.3. 
The simulation parameters are summarized in Table~\ref{RUNTABLE}. 
The number of configurations in our study is
rather small, and we have made no attempt to estimate finite (four dimensional) 
volume systematic errors.  These deficiencies will, of course, be addressed in 
future works.  The lattices correspond to $(1.5\,{\rm fm})^3$ for
$\beta=6.0$ and 6.3 and $(2.1\,{\rm fm})^3$ for 5.85, which 
from previous lattice studies
may result in a deviation of a few percent from the infinite volume case. 
We have begun to address systematic corrections due to finite $N_s$.
All of the correlation functions discussed below were calculated in the
lattice Landau gauge
which was chosen for convenience, in principle any gauge 
will do.

We begin by discussing the numerical investigation of 
Eq.~\ref{cwi} for $O=J_5$. 
First, Eq.~\ref{cwi} is satisfied exactly on any configuration
since it is derived from the corresponding operator identity. We
checked this explicity in our simulations.
In the asymptotic large time limit we get
\BEA
2\sinh{(a m_\pi/2)}
{\LL A_\mu|\pi\RR}/{\LL J_5|\pi \RR}&=&
2 m +2{\LL J_{5q}|\pi\RR}/{\LL J_5|\pi\RR},
\label{ratio}
\EEA
which goes over to the continuum relation for $a m_\pi \ll 1 $ and 
$N_s\to\infty$.
At $\beta=6.0$ and $N_s=10$ we find the l.h.s. of Eq.~\ref{ratio} to be
0.1578(2) and 0.1083(3) for $m=0.075$ and 0.05, respectively. The 
anomalous contributions for these two masses are $2\times$(0.00385(5) and
0.00408(12)), which appears to be roughly constant with $m$.
Increasing $N_s$ to 14 at $m=0.05$, 
the anomalous contribution falls to
$(2\times)$ 0.00152(8) while the l.h.s. is 0.1026(6), 
which shows that increasing $N_s$ really does take us
towards the chiral limit.

\begin{figure}[hbt]
    \vbox{ \epsfxsize=6.0in \epsfbox[0 0 4096 4096]{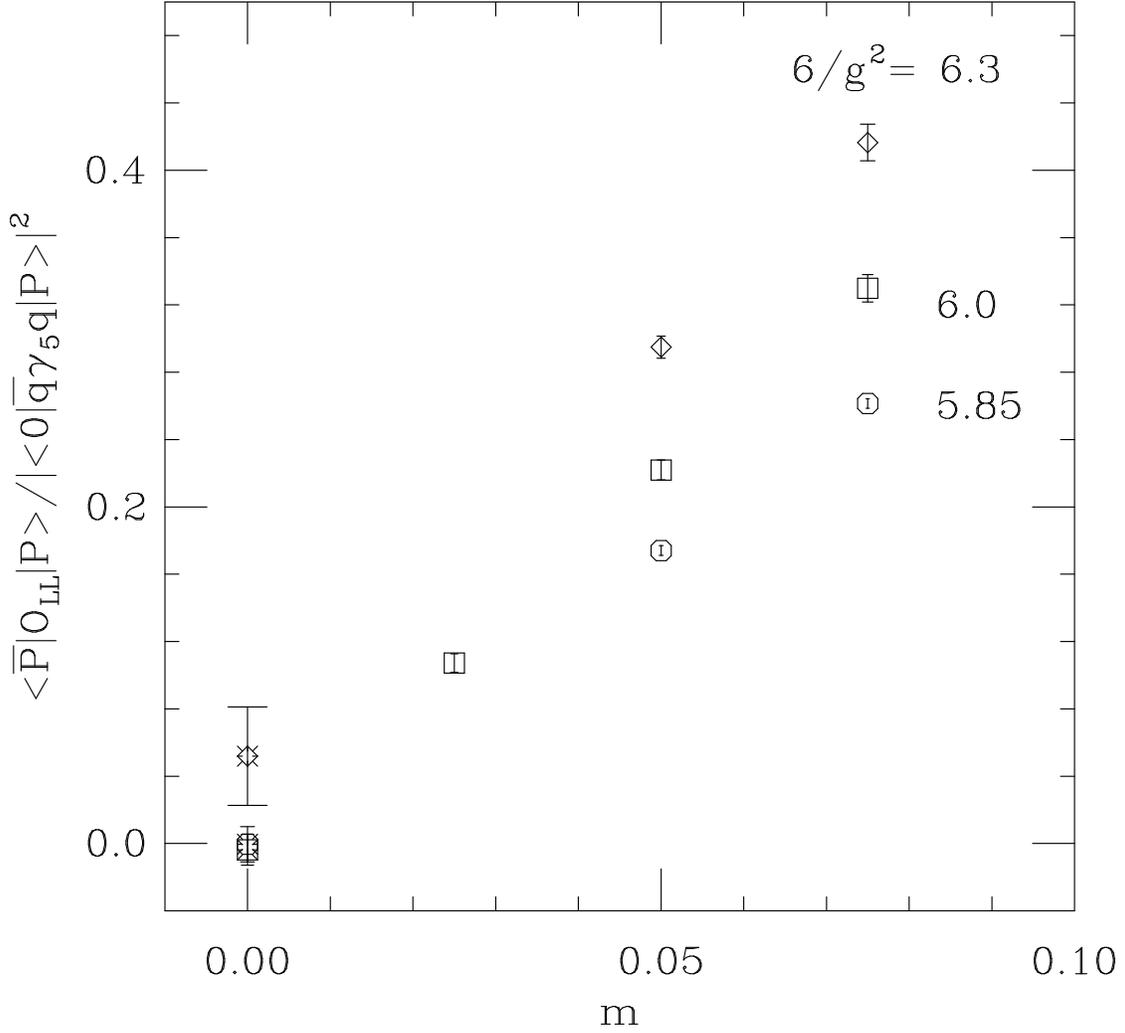} }
    \caption{The matrix element $\LL \bar P |O_{LL}| P \RR$ \vs $m$.
	$\left.|P\RR$ is a non-singlet pseudoscalar state.
	$m$ is proportional to the quark mass in lattice units.
	At $\beta=5.85$ and 6.0 $\MLL$ vanishes linearly with $m$.
	The slight overshoot at $\beta=6.3$ is likely due to higher
	order terms in the chiral expansion(see text). }
    \label{mll}
\end{figure}

While we have not investigated the CWI for $O_{LL}$, the matrix
element $\LL K^0 |O_{LL}|\bar K^0\RR$ vanishes linearly
with $m$ in the chiral limit as required by chiral perturbation
theory and shown in Fig.~\ref{mll}. This indicates that the
anomalous term in Eq.~\ref{cwi} for $O_{LL}$ is highly suppressed.
At $\beta=5.85$ the two
data points extrapolate linearly to -0.0005(100) at $m=0$. At $\beta=6.0$
the three data points extrapolate to -.004(9) with a $\chi^2$ per degree
of freedom of 0.2 for a non-covariant fit.
We do not have enough data to perform a 
covariant fit. At $\beta=6.3$, the two points extrapolate to 
0.05(3). This slight overshoot is not unexpected since the values of
$m$ used are for rather heavy quarks. In our initial study
we found a similar behavior\cite{US}, and 
as the quark mass was lowered, the required linear behavior set in.
Since we do not have a smaller mass at this coupling, there will be
a small systematic increase in $B_K$ since the fit will overestimate 
the matrix element at the strange quark mass.
All of the above results are for $N_s=10$ except at $\beta=5.85$ where
$N_s=14$ was used for reasons explained below.

\begin{figure}[hbt]
    \vbox{ \epsfxsize=6.0in \epsfbox[0 0 4096 4096]{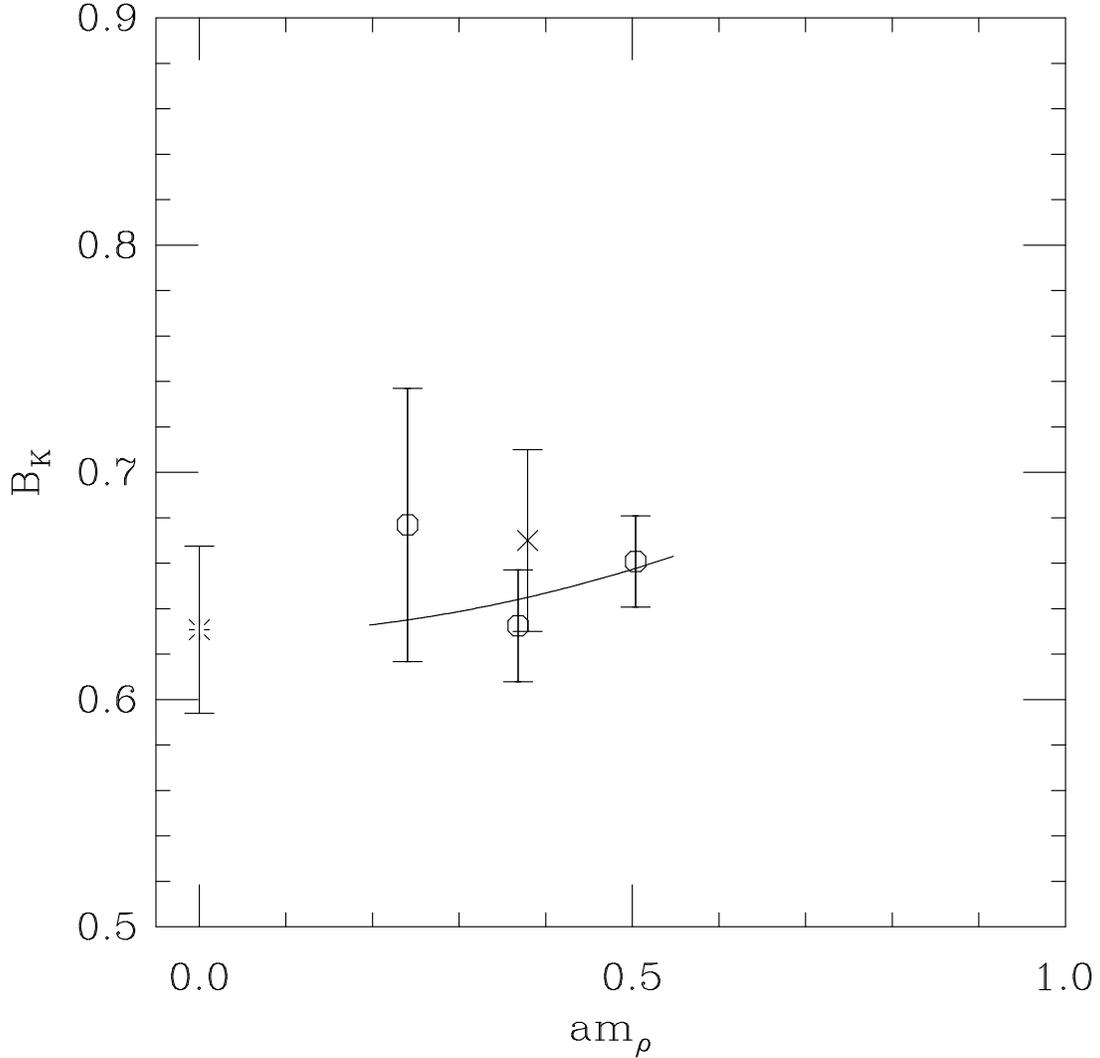} }
    \caption{Kaon B parameter. The
	solid line is a pure quadratic fit to the data, and the burst denotes 
	the extrapolation to
	the continuum limit, $a=0$. The data are for $N_s=10$
	($\beta=6.0$, 6.3) and 14 (5.85). 
	The cross (not used in the fit) denotes
	the partially unquenched result discussed in the text.
	The energy scale is roughly 2 GeV at $\beta=6.0$, and perturbative
	corrections have not been included.}
    \label{BK}
\end{figure}

Fig.~\ref{BK} shows $B_K$  as a function of
$a m_\rho$, or equivalently the lattice spacing. $B_K$ is estimated
at each coupling from a linear fit of the degenerate quark data. The
fit is then evaluated at one-half the value of $m$ 
corresponding to the strange quark as
determined from a fit to the pseudoscalar mass squared (see below). $am_\rho$ is
determined from a simple jackknife average of the effective 
mass over a suitable plateau.
The results for $B_K$ depend weakly on $\beta$, and are well fit to a
pure quadratic in $a$. 
Using this fit, we find $B_K(\mu=a^{-1})=0.628(47)$
in the continuum limit with a confidence level of 0.39. 
This result is already consistent
with the previous Kogut-Susskind result\cite{JLQCD,SHARPE} and 
a recent Wilson quark result\cite{JLQCD2} using CWI's similar to
Eq.~\ref{cwi} to enforce the proper chiral behavior of $O_{LL}$
for Wilson quarks. We note that the data can be fit to a linear
function of the lattice spacing as well, which yields $B_K(\mu=a^{-1})=0.617(80)$, 
though we emphasize again that linear corrections
are expected to be highly suppressed on theoretical grounds. More importantly,
there is no evidence for $O(a)$ corrections in Fig.~\ref{mll}. Similarly,
the denominator in $B_K$, $\MK_{VS}$, exhibits the correct chiral behavor.
Above, 
the notation $B_K(\mu=a^{-1})$
simply means the uncorrected lattice data have been
used to perform the extrapolation;
\ie, our result does not include the
perturbative running of $B_K$ at each lattice spacing to a common energy scale.
This requires a perturbative calculation to determine the renormalization
of $O_{LL}$, which has not yet been done.
The energy scale at $\beta=6.0$,
as determined by the inverse lattice spacing, is roughly 2 GeV.
Since the dependence on
$a$ is already mild, this should not have a significant impact
on our $a=0$ estimate. We also note that the coefficient of the quadratic
term in our fit is 0.12(23), or zero within errors. 
Also, from our previous work\cite{US} which
did not give a value for $B_K$, we find $B_K(\mu=a^{-1})=0.67(4)$ on a set of 20
Kogut-Susskind lattices with $m_{KS}=0.01$ and $\beta=5.7$\cite{COL} and the
same five dimensional lattice volume as the point at $\beta=6.0$. The energy
scale is nearly that of $\beta=6.0$, quenched. This partially unquenched
result indicates that the error from quenching may be small, as was found
for $B_K$ using Kogut-Susskind quarks\cite{ISH,KS-UN}.

At $\beta=6.0$,
we have also calculated $B_K$ using the partially conserved axial
current $A^a_\mu(x)$ (and the analogous vector current) at
$m=0.05$ and 0.075.
This point split conserved current requires explicit factors of the gauge links
to be gauge invariant. Alternatively a gauge non-invariant operator
may be defined by omitting the links; the two definitions
become equivalent in the continuum limit.
Results for the gauge non-invariant operators agree within small 
statistical errors with those for operators constructed from Eq.~\ref{quarks}. 
The results for the gauge invariant operators
are somewhat larger: $B_K^{inv}(\mu=a^{-1})=0.872(22)$ and 0.926(19)
at $m=0.05$ and 0.075, respectively. A similar situation holds in the
Kogut-Susskind case where it was shown that the gauge invariant operators
receive appreciable perturbative corrections which bring the two results
into agreement~\cite{ISH}.

\begin{figure}[hbt]
    \vbox{ \epsfxsize=6.0in \epsfbox[0 0 4096 4096]{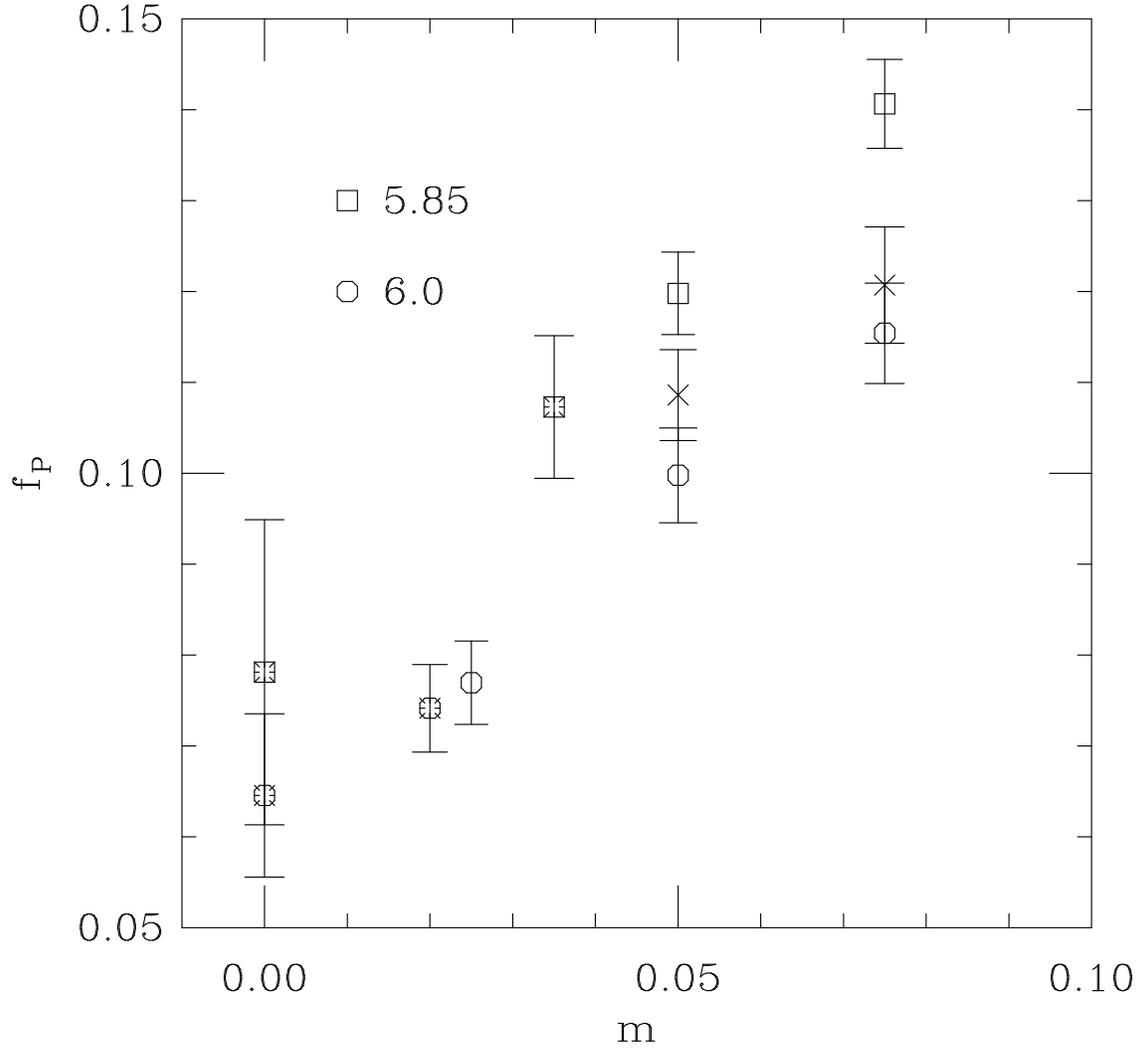} }
    \caption{Pseudoscalar decay constant. Bursts are
	linear extrapolations to values of $m$ corresponding to the
	pion and the kaon. $N_s=10$ and 14 for $\beta=6.0$ and $5.85$,
	respectively. The crosses denote values calculated from the
	matrix element of the partially conserved axial current
	($\beta=6.0$ only).} 
    \label{decay}
\end{figure}

Using Eq.~\ref{ratio}, neglecting the anomalous contribution, and
using the definition of the decay constant
$f_P m_P\equiv \LL 0| A^a_0|P\RR$,
we can determine the pseudoscalar decay constant from the measurement of 
$\LL 0| J^a_5|P\RR$. Performing simultaneous covariant fits to the wall-point
and wall-wall correlators of $J_5$ yields the matrix element. At $\beta=5.85$
and 6.0 each fit has a good confidence level (CL $\approx 0.7$).
The results are shown in Fig.~\ref{decay}. 
Proceeding as before with $B_K$, we find $f_K=159(14)$ MeV 
and 164(12) MeV for $\beta=6.0$ and 5.85, respectively. 
The errors are statistical and do not include 
the error in the lattice spacing determination from $am_\rho$.
The central values agree with experiment,
$f_{K^+}= 160$ MeV.
The lattice spacing determinations from $am_\rho$ 
give $a^{-1}=1.53(27)$, 2.09(21),
and 3.20(81) GeV at $\beta=5.85$, 6.0, and 6.3, respectively. These are similar
to Wilson and Kogut-Susskind lattice spacings for similar quenched lattices. 
Alternatively, we may form the dimensionless ratio $f_K/m_\rho$. We find
for $\beta=5.85$ and 6.0, $f_K/m_\rho=0.213(42)$ and 0.206(27), 
where we have added the statistical errors naively in quadrature. The
experimental result is 0.208. At present our data
at $\beta=6.3$ are too noisy after extrapolation to give a significant
result. Finally we note that we have also calculated the decay constant directly
from the matrix element of the partially conserved axial current 
at the points $m=0.05$ and 0.075 at $\beta=6.0$, and the results agree with
those using the matrix element of the pseudoscalar density
(see Fig.~\ref{decay}). 
While the above results indicate good scaling behavior, they
must be checked further with improved statistics and a fully
covariant fitting procedure. 
More importantly, the continuum limit still has to be taken:
a recent precise calculation using quenched Wilson
quarks by the CP-PACS collaboration gives a value for $f_K/m_\rho$ in
the continuum limit that is inconsistent with experiment~\cite{CPPACS},
so the above agreement with experiment may be fortuitous. 

\begin{figure}[hbt]
    \vbox{ \epsfxsize=6.0in \epsfbox[0 0 4096 4096]{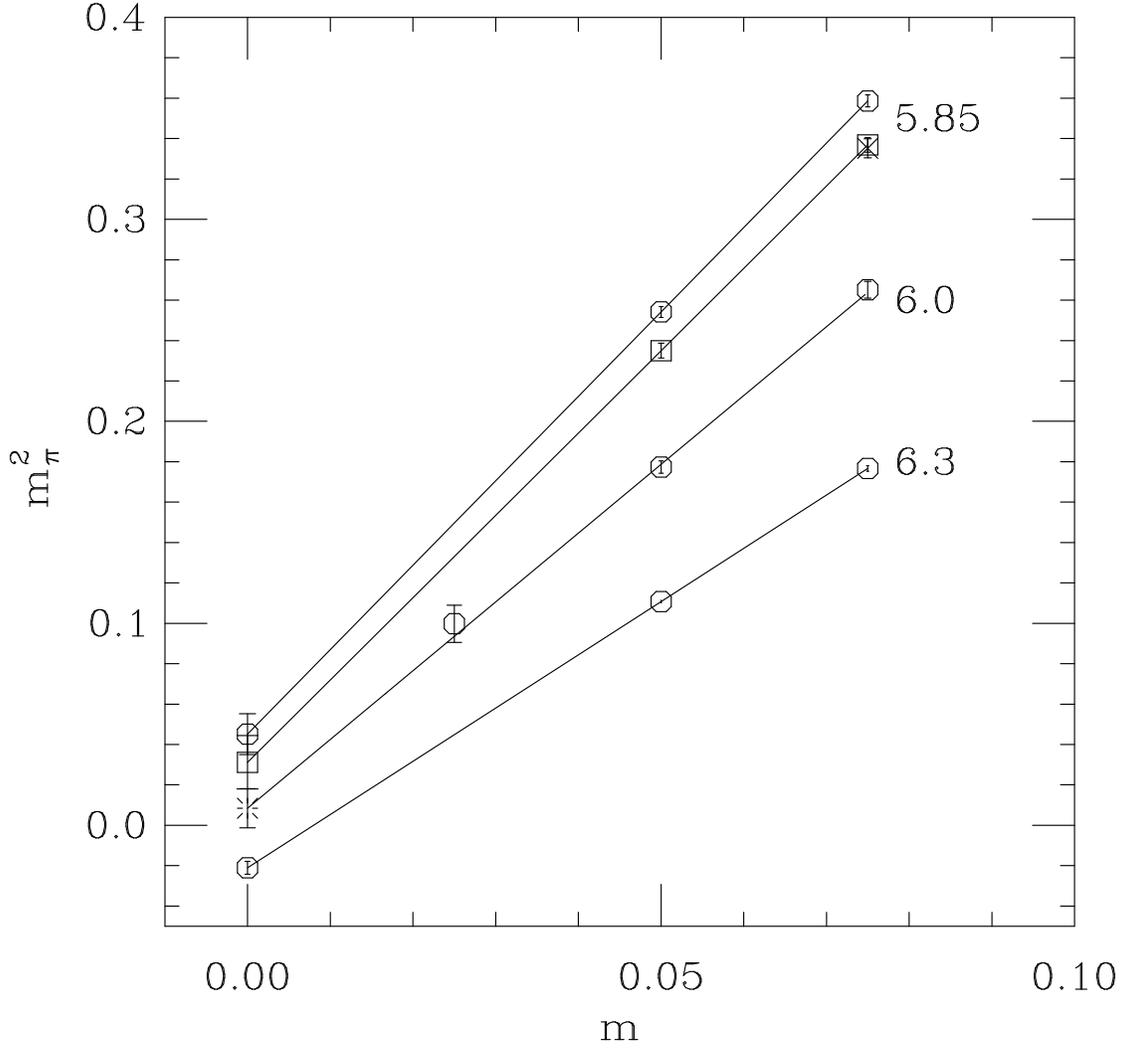} }
    \caption{Pion mass squared. The curve at $\beta=6.0$ 
	($N_s=10$(octagons)) agrees with the expectation from lowest
	order chiral perturbation theory.
	The others show small but significant deviations.
	At $\beta=5.85$ going from  $N_s=10$ to 14 reduces the discrepancy. 
	However increasing $N_s$ to 18(cross) has no effect. 
	At $\beta=6.3$, the discrepancy
	is most likely due to higher order terms in the 
	chiral expansion (see text).}
    \label{mpi2}
\end{figure}

In Fig.~\ref{mpi2} we show the pion mass squared as a function of
$m$. Lowest order chiral perturbation theory requires $m_\pi^2$ to
vanish linearly with $m$. 
At $\beta=6.0$ our three data points extrapolate linearly to .008(10), which
agrees with the above expectations. At $\beta=6.3$, the two data points
extrapolate to $-0.021(3)$, which is once again likely due to
quark masses that are too heavy to agree with lowest order chiral perturbation
theory. At $\beta=5.85$ the two masses extrapolate to 0.045(10) for $N_s=10$
and 0.031(13) for $N_s=14$. This discrepancy is probably not due to higher order
terms in the chiral expansion since the physical quark masses are 
light compared to 
the masses at the other couplings, and the curvature would have the wrong sign.
We see a large downward shift in $m_\pi^2$ as $N_s$ goes from 10 to 14. However,
increasing $N_s$ to 18 at $m=0.075$ has a negligible effect. This behavior may
signal a strong coupling effect where the suppression of explicit 
chiral symmetry breaking terms with $N_s$ may be weakened. 
In the case of the vector Schwinger model, it was
found that topology changing gauge configurations can induce 
significant explicit chiral symmetry breaking effects\cite{VRANAS}. 
Further investigation is required.


Our study shows that DWF are an attractive alternative to Kogut-Susskind and
Wilson quarks for lattice QCD calculations where chiral symmetry is crucial.
For weak matrix elements in particular, DWF yield good agreement with 
expectations from chiral perturbation theory without the 
complicated mixing of operators required with
Wilson quarks, or the entanglement of flavor and space-time 
degrees of freedom as with Kogut-Susskind quarks. 
Perhaps even more importantly, 
up to exponentially small corrections,
DWF maintain the full chiral symmetry of QCD 
at relatively strong couplings, and thus
should have more continuum-like behavior.
The data presented here seem to indicate just that, though future studies
with improved statistics are needed to confirm this. 
This improved scaling may
compensate for the added cost of the extra dimension.

Our domain wall fermion code relies heavily on the four dimensional MILC
code\cite{MILCCODE}, which we are happy to acknowledge again.
This research was supported by US DOE grant
DE-AC0276CH0016. The numerical computations were carried out on the 
NERSC T3E.


\begin{thebibliography}{-29}
\bibitem{US}{T. Blum and A. Soni, Phys. Rev. {\bf D56}, 174 (1997).}
\bibitem{KAPLAN}{D. Kaplan, Phys. Lett. B {\bf 288}, 342 (1992).}
\bibitem{SHAMIR}{Y. Shamir, Nucl. Phys. B {\bf 406}, 90 (1993).}
\bibitem{VRANAS} P.~M.~Vranas, hep-lat/9705023.
\bibitem{SHAMIRandFURMAN}{Y. Shamir and V. Furman, Nucl. Phys. B {\bf 439}, 54 (1995).}
\bibitem{JLQCD}S. Aoki, \etal, Nucl. Phys. B (Proc.Suppl. {\bf 53}), 341 (1997).
\bibitem{SHARPE}S. Sharpe, Nucl. Phys. B(Proc.Suppl. {\bf 34}) , 403 (1994).
\bibitem{JLQCD2}S. Aoki, \etal, hep-lat/9705035.
\bibitem{COL}{F. Brown, \etal, Phys. Rev. Lett. {\bf 67}, 1062 (1991).}
\bibitem{ISH}N. Ishizuka, \etal, Phys. Rev. Lett. {\bf 71}, 24 (1993);
\bibitem{KS-UN} G. Kilcup, Phys. Rev. Lett. {\bf 71}, 1677 (1993); 
W. Lee and M. Klomfass, hep-lat/9608089.
\bibitem{CPPACS}See the plenary talk by T. Yoshi\'e at Lattice 97.
\bibitem{MILCCODE}{C.~Bernard, \etal,
in {\it Workshop on Fermion Algorithms}, edited by H.~J.~Hermann and
F.~Karsch, (World Scientific, Singapore, 1991). }
\end{thebibliography}
\end{document}